\documentclass[aps,prl,twocolumn,showpacs,preprintnumbers,amsmath,amssymb]{revtex4-1}
\usepackage[breaklinks=true,colorlinks=true
,urlcolor=blue
,anchorcolor=blue
,citecolor=blue
,filecolor=blue
,linkcolor=blue
,menucolor=blue
,pagecolor=blue
,linktocpage=true
,pdfproducer=medialab
,pdfa=true
]{hyperref}
\usepackage{graphicx,hyperref,color}
\usepackage{color}
\usepackage[dvipsnames]{xcolor}
\usepackage{graphicx}
\usepackage{dcolumn}
\usepackage{bm}
\usepackage{bbm}
\usepackage{mathtools}

\usepackage{tikz}
\usetikzlibrary{arrows,arrows.meta,intersections, calc,positioning,decorations.pathreplacing,decorations.pathmorphing,shapes}
\usetikzlibrary{patterns}
\usetikzlibrary{decorations.markings}
\usetikzlibrary{knots}

\def\ep{{\epsilon}}

\def\frac#1#2{{#1\over #2}}

\def\G{{\Gamma}}

\def\g{{\gamma}}
\def\s{\sqrt}

\def\be{\begin{equation}}
\def\ee{\end{equation}}
\def\ba{\begin{eqnarray}}
\def\ea{\end{eqnarray}}
\def\de{\partial}

\def\ddd{\cdot\cdot\cdot}
\def\no{\nonumber \\}

\def\ep{\epsilon}

\newcommand{\Sdis}{S_{\mathrm{dis}}}
\newcommand{\Sint}{S_{\mathrm{int}}}
\newcommand{\Scon}{S_{\mathrm{con}}}

\usepackage{braket}

\begin{document}

\title{{\large \textbf{The $g$-theorem from Strong Subadditivity}}}
\preprint{YITP-24-32}

\author{Jonathan Harper$^a$, Hiroki Kanda$^{a}$,  Tadashi Takayanagi$^{a,b}$ and Kenya Tasuki$^a$}

\affiliation{$^a$Center for Gravitational Physics, Yukawa Institute for Theoretical Physics, Kyoto University, \\
Kitashirakawa Oiwakecho, Sakyo-ku, Kyoto 606-8502, Japan}

\affiliation{$^b$Inamori Research Institute for Science, 620 Suiginya-cho, Shimogyo-ku, Kyoto 600-8411, Japan}


\begin{abstract}
We show that strong subadditivity provides a simple derivation of the $g$-theorem for the boundary renormalization group flow in two-dimensional conformal field theories. We work out its holographic interpretation and also give a derivation of the $g$-theorem for the case of an interface in two-dimensional conformal field theories.
We also geometrically confirm strong subadditivity for holographic duals of conformal field theories on manifolds with boundaries. 
\end{abstract}

\maketitle



{\bf 1. Introduction}

Strong subadditivity (SSA)~\cite{Lieb:1973cp,Lieb:1973zz}
\be
S_{AB}+S_{BC}-S_{ABC}-S_{B}\geq 0,  \label{ssar}
\ee
is a fundamental property which explains the nature of quantum information in the form of certain monotonicity relation, analogous to the second law of thermodynamics.
For example, SSA shows that the conditional mutual information is non-negative. Here we write the entanglement entropy for the subsystem $A$ as $S_A$. 
To define $S_A$ we introduce the reduced density matrix $\rho_A$ by tracing the density matrix for the whole system over the complement of the region $A$ and then consider its von Neumann entropy $S_A=-\mbox{Tr}\rho_A\log \rho_A$.

SSA also plays an important role in quantum field theories (QFTs) as it offers a universal property for the degrees of freedom under the renormalization group (RG) flow. 
Indeed we can derive the $c$-theorem~\cite{Casini:2004bw} in two-dimensional (2d) QFTs and the $F$-theorem~\cite{Casini:2012ei} in 3d QFTs from the SSA relation (\ref{ssar}). 
The $a$-theorem in 4d QFTs was shown via a more elaborate method in~\cite{Casini:2017vbe}. 

\begin{figure}[ttt]
   \centering
   \includegraphics[width=4cm]{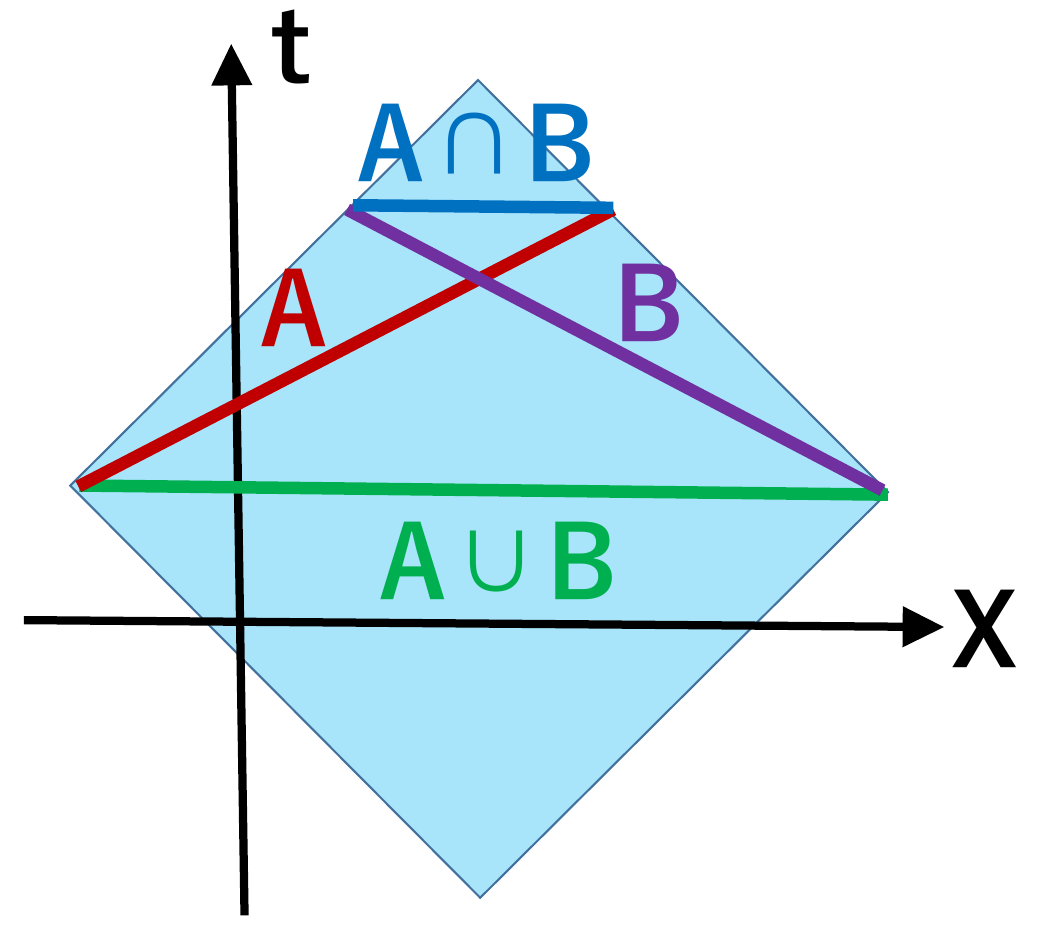}
   \caption{The setup of deriving entropic $c$-theorem.}
   \label{fig:DA}
\end{figure}

Let us briefly recount the entropic $c$-theorem in the 2d case~\cite{Casini:2004bw}. 
Consider the entanglement entropy $S_A$ for an interval $A$. 
We write its Lorentz invariant length as $|A|=l$, and then the entropy becomes a function of $l$, which is expressed as $S_A(l)$. 
It is also useful to rewrite SSA (\ref{ssar}) as
\be
S_{A}+S_{B}\geq S_{A\cup B}+S_{A\cap B},  \label{ssaeq}
\ee
where we regard $AB$ and $BC$ in (\ref{ssar}) as $A$ and $B$, respectively. 
By taking advantage of the relativistic invariance of 2d QFT, we can choose the subsystems $A,B,A\cap B$ and $A\cup B$ as in Fig.~\ref{fig:DA}. 
If we set $|A\cap B|=l_1$ and $|A\cup B|=l_2$, then we find $|A|=|B|=\s{l_1l_2}$.
Thus, SSA (\ref{ssaeq}) leads to the inequality $2S_A(\s{l_1l_2})\geq S_A(l_1)+S_A(l_2)$, which implies that $S_A$ is concave as a function of $\log l$:
\be
\frac{d}{dl}\left[l\frac{dS_A(l)}{dl}\right]\leq 0.
\label{entropc}
\ee
The entanglement entropy for 2d CFT vacua is known to take the form $S_A=(c/3) \log(l/\ep)$, where $c$ is the central charge and $\ep$ is the UV cutoff~\cite{Holzhey:1994we}. 
Therefore, we can regard $C(l)=3l\frac{dS_A(l)}{dl}$ as an effective central charge at the length scale $l$. 
In this way, the inequality (\ref{entropc}) shows the $c$-theorem, which states that the degrees of freedom monotonically decrease under the RG flow. 

Even though the $c$-theorem was originally derived using the more traditional field-theoretic method~\cite{Zamolodchikov:1986gt}, the above SSA argument provides us with a much simpler derivation and shows that at its essence lies the monotonicity of quantum information.

The purpose of this letter is to extend this beautiful and geometrical derivation of the important monotonicity of QFTs, using the entanglement entropy, to cases with boundaries or defects when their bulk theories are conformally invariant.\\

{\bf 2. Entropic derivation of the $g$-theorem for BCFTs}

Consider a 2d CFT on a 2d Lorentzian flat spacetime, whose coordinates are denoted by $(x,t)$ and put a time-like boundary at $x=0$ by limiting the spacetime to the right half plane $x\geq 0$. 
When the boundary condition at $x=0$ preserves a half of the bulk conformal invariance, this theory is called a boundary conformal field theory (BCFT)~\cite{Cardy:1984bb}. 

It is known that the entanglement entropy for an interval $A$ which stretches from the boundary $x=0$ to a point $x=\xi$ at any time $t=t_0$, takes the form~\cite{Calabrese:2004eu}:
\ba
S_A=\frac{c}{6}\log\frac{2\xi}{\ep}+\log g,  \label{crfors}
\ea
where $\ep$ is the UV cutoff and $\log g$ is called the boundary entropy.

\begin{figure}[ttt]
   \centering
   \includegraphics[width=5cm]{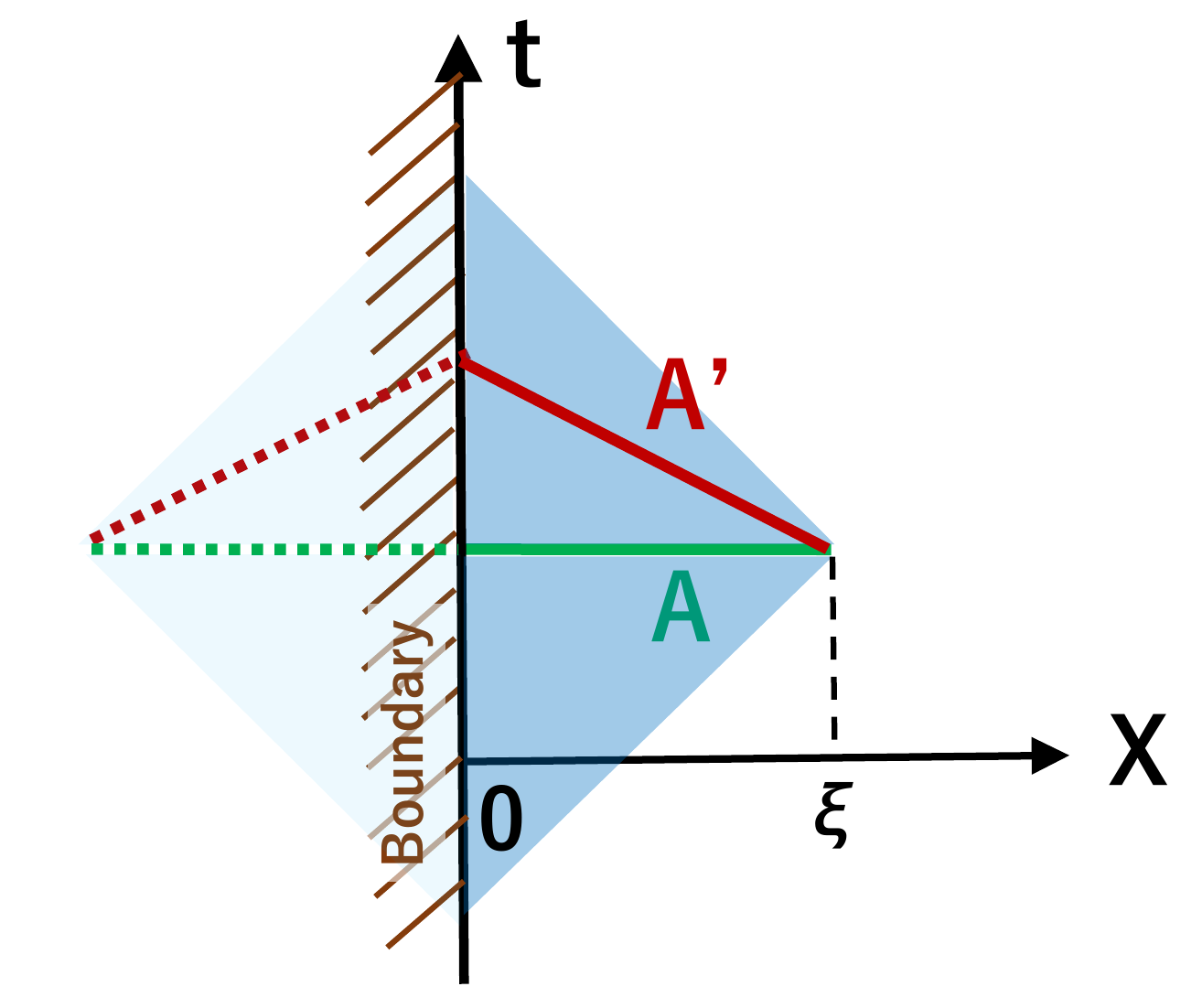}
   \caption{Sketches of subsystem $A$ and $A'$ entanglement entropy in a 2d BCFT on a half plane $x>0$. Since they have the same domain of dependence (blue region), we find $S_A=S_{A'}$.}
   \label{fig:DAB}
\end{figure}

Even if we deform the subsystem (called $A'$) such that it ends on $(\xi,t_0)$ and a boundary point $x=0$ at a time $t_0-\xi<t<t_0+\xi$, which is within the domain of dependence of $A$ (and its mirror), the entanglement entropy does not change, i.e., $S_A=S_{A'}$. 
See Fig.~\ref{fig:DAB} for a sketch. 
This is true for any relativistic field theory with a boundary and is due to the complete reflection at the boundary.

Now, we break the conformal invariance at the boundary by a relevant boundary perturbation $\int dt \, O(t,x=0)$.
The basic property that the degrees of freedom at the boundary monotonically decrease under the boundary RG flow is known as the $g$-theorem~\cite{Affleck:1991tk}. 
The $g$-theorem argues that the boundary entropy $\log g$ in (\ref{crfors}) as a function of length scale, so-called the $g$-function, is monotonically decreasing under the boundary RG flow. 
This theorem was proved by using the boundary RG flow in~\cite{Friedan:2003yc} and by using the relative entropy in~\cite{Casini:2016fgb,Casini:2022bsu}. 
For higher dimensional versions of g-functions, refer to~\cite{Nozaki:2012qd,Jensen:2013lxa,Jensen:2015swa,Kobayashi:2018lil,Casini:2018nym,Jensen:2018rxu,Goto:2020per,Wang:2020xkc,Nishioka:2021uef,Wang:2021mdq,Cuomo:2021rkm,Shachar:2022fqk}.

\begin{figure}[ttt]
   \centering
   \includegraphics[width=7cm]{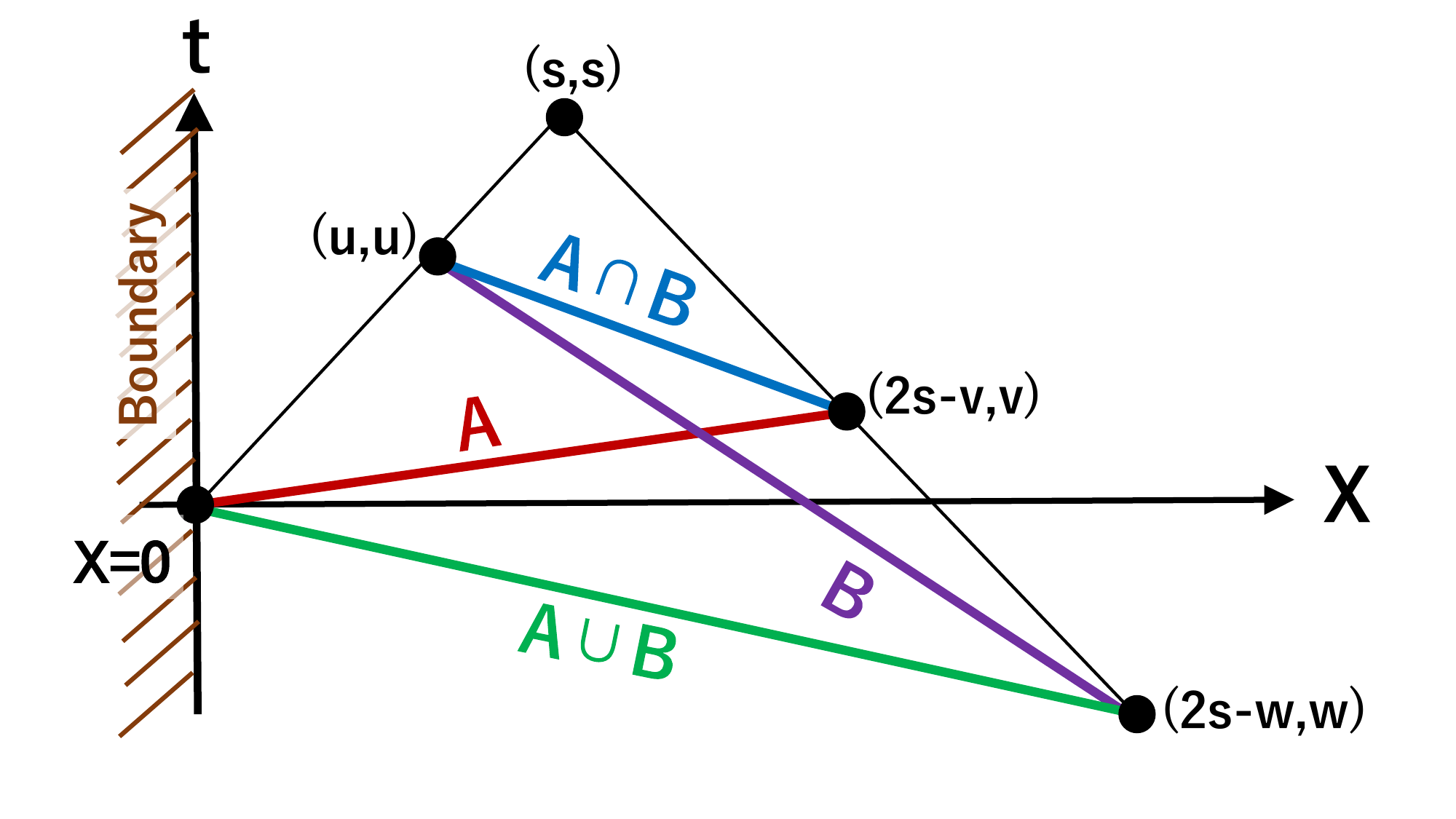}
   \caption{The Lorentzian setup for the SSA in a 2d BCFT.}
   \label{HEESetup}
\end{figure}

Below, we would like to present another simpler derivation of the $g$-theorem directly from SSA. 
Consider the Lorentzian setup of Fig.~\ref{HEESetup} and the implication of SSA:
\ba
\varDelta{S} \coloneqq S_{A}+S_{B}-S_{A\cup B}-S_{A\cap B}\geq 0.   \label{dels}
\ea
We write the entanglement entropy $S_A$ for an interval $A$ as $S(x_1,t_1;x_2,t_2)$, whose endpoints are set to be $P_1:(x_1,t_1)$ and $P_2:(x_2,t_2)$. 
When $P_1$ is situated at the boundary $x_1=0$, then the entanglement entropy $S_A$ only depends on $x_2$ as we already explained in Fig.~\ref{fig:DAB}, and we write this as $\Sdis(x)$. 

Now we choose the subsystems such that each of the space-like intervals $A,B,A\cup B$ and $A\cap B$ connects two points on the two null lays which intersect at the point $(s,s)$ and such that they satisfy $|A||B|=|A\cup B||A\cap B|$, as illustrated in Fig.~\ref{HEESetup}. 
Then, their entanglement entropies are described by 
\ba
&& S_{A\cup B}=\Sdis(2s-w),\ \ S_{A\cap B}=S(u,u;2s-v,v),\no
&& S_A=\Sdis(2s-v),\ \ S_B=S(u,u;2s-w,w),\label{eedefg}
\ea
where we assume $w<v<s$ and $s>0$. 
Below, we appropriately choose the values of $s,u,v$ and $w$ to obtain the tightest bound from SSA.

First, we take the limit $u\to s$, where $B$ and $A\cap B$ become light-like, which is equivalent to the zero width or equally the UV limit. We can understand this by regarding the two point function of twist operators which compute the entanglement entropy as a four point function via the mirror method, which is factoring into a square of two point functions of null separated twist operators. Moreover, this claim is also obvious in the holographic dual of BCFTs \cite{Takayanagi:2011zk,Fujita:2011fp,Karch:2000gx},
where the extremal surface dual to $S_A$ is localized near the boundary. 

Therefore, in this limit, we can approximate $S_B$ and $S_{A\cap B}$ by their values in the CFT vacuum ignoring the presence of the boundary at $x=0$:
\ba
 S_B\!&\simeq& \!\frac{c}{3}\log\frac{|B|}{\ep}=\frac{c}{6}\log \left[4(s-u)(s-w)/\ep^2\right],\no
S_{A\cap B}\!&\simeq&\! \frac{c}{3}\!\log\frac{|A\cap B|}{\ep}\!=\!\frac{c}{6}\log\! \left[4(s-u)(s-v)/\ep^2\right]\!.  \label{limitee}
\ea
Thus $\varDelta{S}$ defined in (\ref{dels}) is evaluated to be
\ba
\varDelta{S} =\Sdis(2s-v)-\Sdis(2s-w)+\frac{c}{6}\log\frac{s-w}{s-v}.  \label{ineqssa}
\ea

Next, we take the value of $v$ very close to $w$ by setting $v=w+\delta$, where $\delta$ is an infinitesimally small and positive constant. 
Then (\ref{ineqssa}) can be rewritten as 
\ba
\varDelta{S}=\delta\cdot \left(-\frac{d\Sdis(\xi)}{d\xi}\Biggl|_{\xi=2s-w}+\frac{c}{6}\cdot\frac{1}{s-w}\right).
\ea

Finally, by assuming $w<0$ and taking $s$ to be very small such that $s\ll |w|$, we find that the SSA $\varDelta{S} \geq 0$ gives the tightest bound:
\ba
\xi\frac{d\Sdis(\xi)}{d\xi}\leq \frac{c}{6},  \label{strongin}
\ea
where $\xi\simeq -w>0$ takes an arbitrary positive value.

Now we define the entropic $g$-function $g(\xi)$ at the length scale $x$ by
\ba
\log g(\xi)\coloneqq \Sdis(\xi)-\frac{c}{6}\log \frac{2\xi}{\ep},
\ea
such that it gives the boundary entropy at each fixed point following the formula (\ref{crfors}). 
Then SSA (\ref{strongin}) leads to the inequality:
\ba
\frac{d}{d\xi}\log g(\xi)\leq 0.
\ea
This completes the derivation of the entropic $g$-theorem.\\

\begin{figure}[ttt]
   \centering
   \includegraphics[width=8cm]{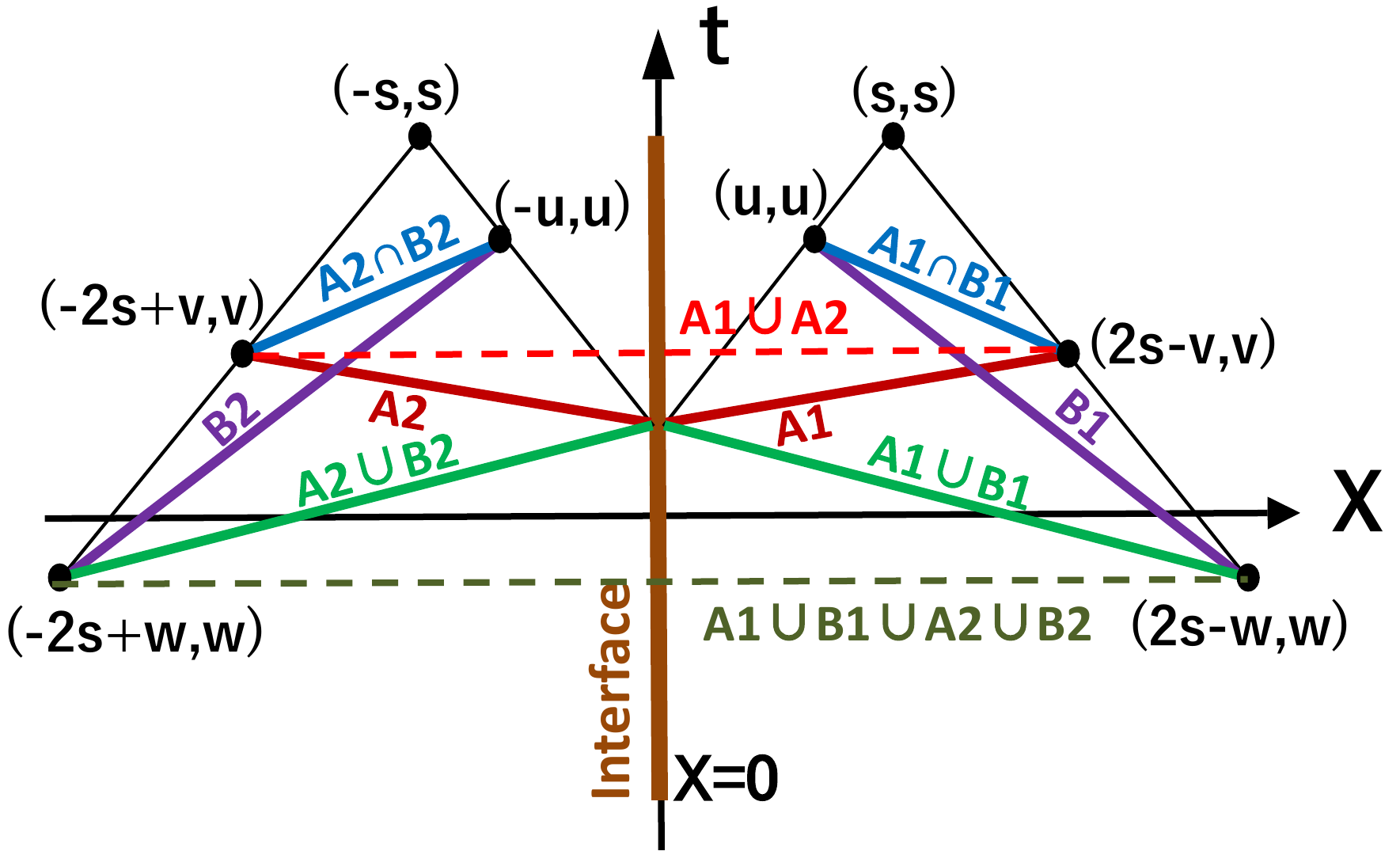}
   \caption{The Lorentzian setup for the SSA in a 2d ICFT. The horizontal dotted intervals provide the $g$-function.}
   \label{fig:HEEdefectSetup}
\end{figure}

{\bf 3. Entropic $g$-theorem for interface CFTs}
Next, we extend our previous derivation of the $g$-theorem to interfaces in 2d CFTs. 
Consider a 2d CFT with central charge $c$ on the $(x,t)$ plane and place an interface along $x=0$ as depicted in Fig.~\ref{fig:HEEdefectSetup}. 
If the interface preserves half of the bulk conformal symmetry, a so-called interface CFT \cite{Oshikawa:1996ww,Bachas:2001vj,Sakai:2008tt}, then the entanglement entropy $\Sint(\xi)$ for the interval $-\xi\leq x\leq \xi$ at any time $t_0$ takes the form
\ba
\Sint(\xi)=\frac{c}{3}\log\frac{2\xi}{\ep}+\log g_I,
\ea
where the constant $\log g_I$ is the interface entropy. 
When we consider a relevant perturbation localized on the interface, the entropic $g$-theorem for interface CFTs claims that the entropic g-function
\ba
\log g_I(\xi)=\Sint(\xi)-\frac{c}{3}\log\frac{2\xi}{\ep},
\ea
is monotonically decreasing as a function of $\xi$. Refer to~\cite{Azeyanagi:2007qj} for an earlier attempt toward this type of entropic $g$-theorem. 
Below, we will give a complete derivation from SSA. 
Refer also to~\cite{Karch:2023evr} for an interesting implication from SSA when the interface connects two different CFTs with different central charges. 

Our argument goes in parallel with our previous one in BCFTs. 
By doubling the setup of Fig.~\ref{fig:DAB}, we choose the subsystems depicted in Fig.~\ref{fig:HEEdefectSetup}. 
We have two copies of the boosted subsystems, each of which is identical to the ones $A,B,\ddd$ in  Fig.~\ref{fig:DAB}, named as $A_1,B_1,\ddd$ and $A_2,B_2,\ddd$. 
Then we set $A=A_1\cup A_2$ and $B=B_1\cup B_2$ in the SSA relation (\ref{dels}). 
In the $u\to s$ limit, this inequality leads to 
\ba
\varDelta{S} =\Sint(2s-v)-\Sint(2s-w)+\frac{c}{3}\log\frac{s-w}{s-v},
\ea
which is a straightforward extension of (\ref{ineqssa}). 
As in the BCFT case, we further consider the limit $v\to w$ and $s\ll |w|$, and we finally obtain 
\ba
\xi\frac{d\Sint(\xi)}{d\xi}\leq \frac{c}{3},
\ea
which is equivalent to the $g$-theorem $\frac{d\log g_I(\xi)}{d\xi}\leq 0$.\\

{\bf 4. Holographic SSA and the Null Energy Condition}

The anti-de Sitter/conformal field theory (AdS/CFT) correspondence argues that gravity on a $d+1$ dimensional AdS spacetime is equivalent to a $d$ dimensional CFT~\cite{Maldacena:1997re,Gubser:1998bc,Witten:1998qj}. 
In AdS/CFT, we can calculate the entanglement entropy $S_A$ in a geometrical way, known as the holographic entanglement entropy (HEE)~\cite{Ryu:2006bv,Ryu:2006ef,Hubeny:2007xt}. 
It is computed from the area of an extremal surface $\Gamma_A$, denoted by $|\Gamma_A|$, which ends on the boundary of and is homologous to the subsystem $A$ in AdS as 
\ba
S_A=\frac{|\Gamma_A|}{4G_N},  \label{HEE}
\ea
where $G_N$ is the Newton constant in the AdS gravity.
Interestingly, this HEE allows us to derive SSA in a more geometrical way~\cite{Headrick:2007km,Wall:2012uf}, which essentially follows from the triangle inequality in classical geometry.

We can extend the AdS/CFT correspondence to the gravity dual of a CFT on a manifold with boundaries by introducing end-of-the-world-branes (EOW branes)~\cite{Takayanagi:2011zk,Fujita:2011fp,Karch:2000gx}, called the AdS/BCFT correspondence. 
On the EOW brane, we impose the Neumann boundary condition 
\ba
K_{ab}-Kh_{ab}=8\pi G_N T^{(E)}_{ab},
\ea
where $h_{ab},K_{ab}$ and $T^{(E)}_{ab}$ are the induced metric, extrinsic curvature and matter energy stress tensor on the EOW brane. 
The HEE in AdS/BCFT is again given by the formula (\ref{HEE}) with an important addition that the extremal surface $\Gamma_A$ can end on an EOW brane~\cite{Takayanagi:2011zk,Fujita:2011fp}. This can be viewed as a change in the homology constraint such that $\Gamma_A$ is homologous to $A$ \emph{relative} to the EOW brane.

\begin{figure}[ttt]
   \centering
   \includegraphics[width=8cm]{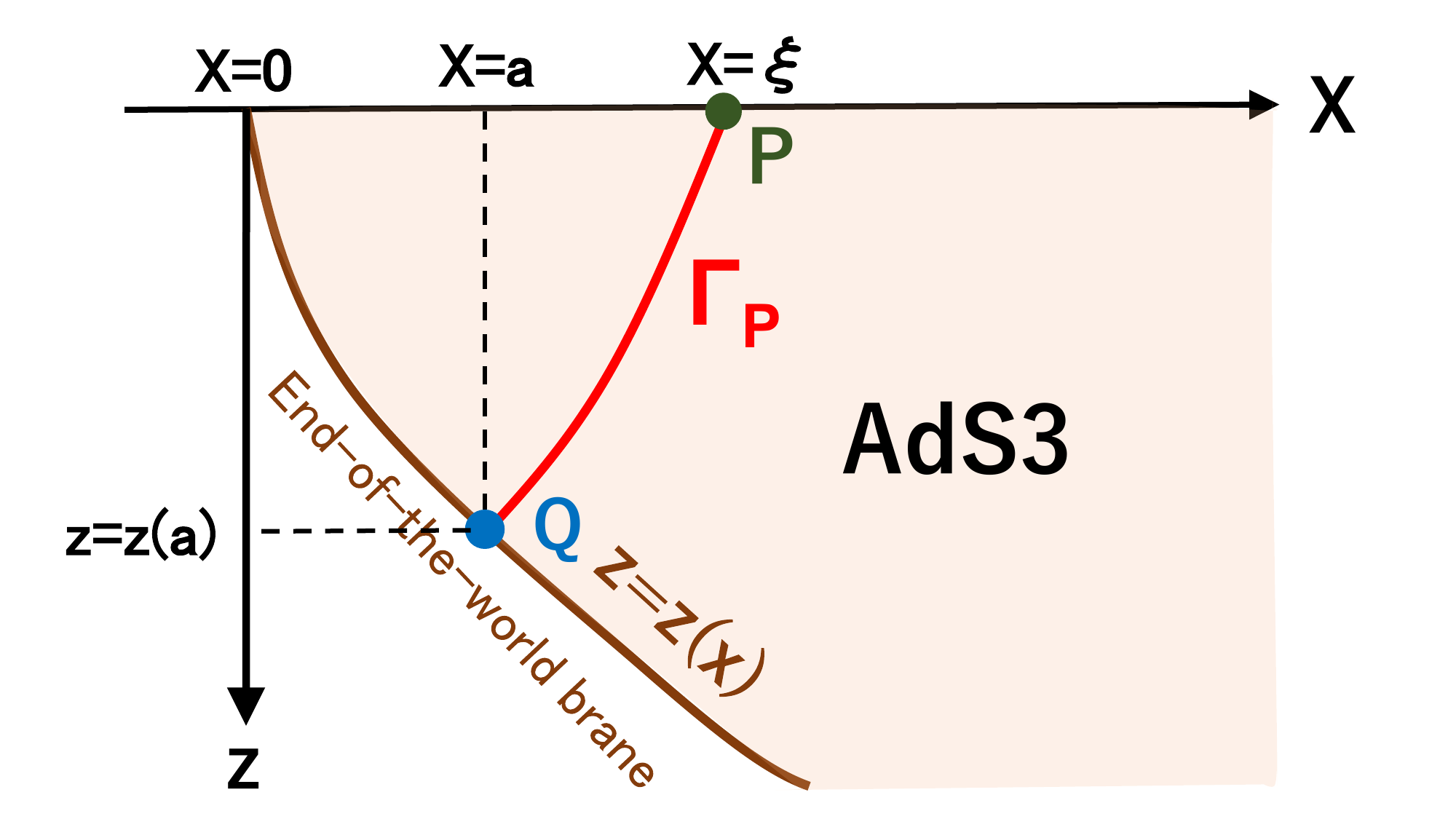}
   \caption{The calculation of geodesic length in AdS$_3/$BCFT$_2$.}
   \label{SimpleSetup}
\end{figure}

For a 2d CFT defined on a space with a boundary, its gravity dual is given by a region of 3d AdS (AdS$_3$) surrounded by an EOW brane. 
Assuming the pure gravity theory in the bulk, we can always choose the metric to be that of the pure AdS$_3$
\ba
ds^2=\frac{dz^2-dt^2+dx^2}{z^2}.
\ea
We specify the profile of EOW brane by $z=z(x)$ such that $z(0)=0$ as in Fig.~\ref{SimpleSetup}, assuming that it is static. 
The gravity dual is given by the region $z<z(x)$. 
The cutoff $\ep$ of the $z$ coordinate is identified with the UV cutoff $\ep$ of the dual CFT. 
When $z(x)\propto x$, the boundary preserves the conformal invariance, i.e., becomes a 2d BCFT. 
In general, the non-trivial profile of $z(x)$ encodes the detailed information of the boundary RG flow (see e.g.\cite{Kanda:2023zse} for an example).

Let us calculate the HEE by using this 3d holographic setup and compare it with our previous arguments for the $g$-theorem.
When the subsystem $A$ is given by an interval which stretches from the boundary $x=0$ to a point $x=\xi$ at any time, the HEE $S_A$ is given by 
\ba
\Sdis(\xi)=\frac{c}{6}|\Gamma_P|. \label{disent}
\ea

For this, let us calculate the length of the geodesic $\Gamma_P$, which connects between a given point $P$ on the AdS boundary $(z,x)=(\ep,\xi)$ and a point $Q$ on the EOW brane, described by $(z,x)=(z(a),a)$. 
The value of $a$ is fixed by minimizing $\Gamma_P$ as a function of $a$. 
Notice that since the EOW brane is static, $\Gamma_P$ is on a constant time slice, leading to $S_A=S_{A'}$ in Fig.~\ref{fig:DAB}.

Since the geodesic $\Gamma_P$ is orthogonal to the EOW brane at $Q$ and is given by a part of a circle in $(x,z)$ plane, we find the relation between $\xi$ and $a$:
\ba
\xi=a-\frac{z(a)}{\dot{z}(a)}+z(a)\s{1+\frac{1}{\dot{z}(a)^2}},
\ea
and the length of the geodesic is computed as
\ba
|\Gamma_P|=\log\left[\frac{2z(a)\s{1+\dot{z}(a)^2}}{\ep(\s{1+\dot{z}(a)^2}+1)}\right].
\ea

For example, if we set $z(x)=\lambda x$, then we find
\ba
\Sdis=\frac{c}{6}|\Gamma_P|=\frac{c}{6}\log\frac{2\xi}{\ep}-\frac{c}{6}\log\left[\frac{1}{\lambda}+\s{1+\frac{1}{\lambda^2}}\right],
\ea
which leads to the standard form of the EE (\ref{crfors}) in 2d BCFT.

For a generic profile $z=z(x)$, we obtain:
\ba
\frac{6}{c}\cdot\xi\frac{\de \Sdis(\xi)}{\de \xi}-1=\frac{az'(a)-z(a)}{z(a)\s{1+z'(a)^2}}. \label{ssader}
\ea
The non-negativity of this quantity is equivalent to the SSA condition (\ref{strongin}). 
Indeed, we can find that (\ref{ssader}) is non-negative if we assume the null energy condition, i.e., $T^{(E)}_{ab}n^an^b\geq 0$ for any null vector $n^a$ in AdS$_3$. 
The null energy condition on the EOW brane leads to the condition $z''(x)\leq 0$ as shown in~\cite{Takayanagi:2011zk,Fujita:2011fp}, where a holographic $g$-theorem was derived. 
This allows us to guarantee $az'(a)-z(a)\leq 0$. 
This is found as follows: 
First, in the UV limit $a\to 0$, we expect the boundary to become conformal, which means $z(a)\propto a$, leading to $az'(a)-z(a)=0$ at $a=0$. 
Moreover, the derivative $(az'(a)-z(a))'=z''(a)$ is non-positive due to the null energy condition. 
Thus, these manifestly show $az'(a)-z(a)\leq 0$. 
In this way, SSA in the setup of Fig.~\ref{fig:DA} precisely requires that the classical gravity satisfies the null energy condition in the gravity dual.\\

{\bf 5. Holographic SSA in Static Backgrounds}

In the above calculations of SSA, it was crucial that we considered the Lorentzian setup taking advantage of boost operations in relativistic QFTs. 
On the other hand, if we assume all subsystems ($A,\ddd$) and the dual extremal surfaces ($\Gamma_A,\ddd$) are on the same time slice $t=t_0$, then we can show that the HEE in AdS/BCFT always satisfies SSA for any profile of the EOW brane at $t=t_0$ (see also~\cite{Chou:2020ffc,Chang:2018pnb} for earlier confirmation of SSA in particular examples). More generally, this claim can also be applied to a setup with a time reversal symmetry $(t-t_0)\to -(t-t_0)$. Note that this does not contradict the null energy condition because we can compensate for the arbitrary shape of the EOW brane at the specific time $t=t_0$ by choosing an appropriate time evolution of the EOW brane profile such that the null energy condition is maintained.

Since the essence of this argument does not depend on the dimension, we will continue to focus on the specific example of AdS$_3/$BCFT$_2$. 
Let $A$ be an interval, whose endpoints are given by $(x_1,t_1)$ and $(x_2,t_2)$, its HEE $S_A$ is computed as the minimum of the area of two configurations of surfaces:
\ba
S_A=\mbox{Min}\left[\Scon(x_1,t_1;x_2,t_2), \Sdis(x_1)+\Sdis(x_2)\right],
\ea
where $\Sdis(x)$ is the disconnected HEE (\ref{disent}) and $\Scon$ is the connected HEE given by
\ba
\Scon(x_1,t_1;x_2,t_2)=\frac{c}{6}\log\left[(x_2-x_1)^2/\ep^2-(t_2-t_1)^2/\ep^2\right].\nonumber
\ea

With these preparations, we can prove SSA for AdS$_3$/BCFT$_2$ on a time slice $\Sigma$ which has the time reversal symmetry.
We take three subsystems $A$, $B$, and $C$ on $\Sigma$. Recalling that the entanglement wedge preserves the order of inclusion, we have
\begin{equation}\label{eq:preserving_inclusion}
    E_B\subset E_{AB},E_{BC}\subset E_{ABC},
\end{equation}
where $E_A$ denotes the homology region which is the entanglement wedge of $A$ projected onto $\Sigma$. This leads to the implication that the intersection $E_{AB}\cap E_{BC}$ is non-empty. Therefore, the extremal surface $\G_{AB}$ of subsystem $AB$ can be divided into three parts:
\begin{equation}
    \G_{AB} = \G_{AB}^{(1)}\cup \G_{AB}^{(2)}\cup\g.
\end{equation}
The first one denotes the part containing $E_{BC}$, i.e. $\G_{AB}^{(1)}:=\G_{AB}\cap E_{BC}$ and the second one is the outer part. The last one is the common part $\g:=\G_{AB}\cap\G_{BC}$. $\G_{BC}^{(1)}$ and $\G_{BC}^{(2)}$ are defined similarly. Furthermore, for subsystem $B$, $Q_B$ denotes the parts of the EOW brane which enclose $E_B$. Namely, by the homology condition on the homology region, we have
\begin{equation}
    \partial E_B = B \cup \G_B \cup Q_B.
\end{equation}

\begin{figure}[htbp]
    \centering
    \includegraphics[width=8cm]{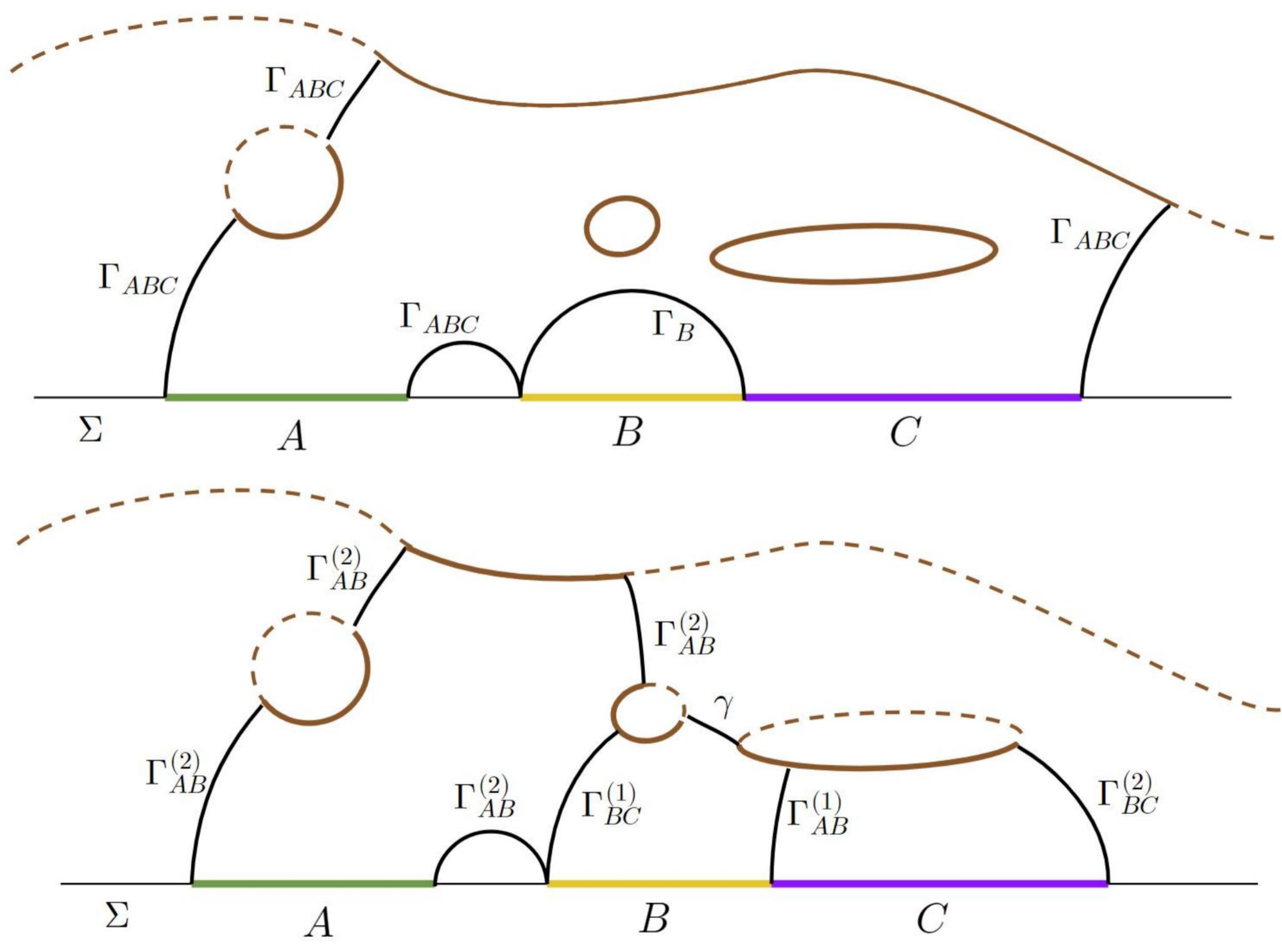}
    \caption{An example of general EOW branes and extremal surfaces. }
    \label{fig:enter-label}
\end{figure}

We define $E^\prime_B$ as the intersection $E_{AB}\cap E_{BC}$, then $E^\prime_B$ satisfies the homology condition:
\begin{equation}
    \partial E_B^\prime = B\cup (\G_{AB}^{(1)}\cup \G_{BC}^{(1)}\cup \g)\cup (Q_{AB}\cap Q_{BC}).
\end{equation}
Because of the extremality of $\G_{B}$, we must have
\begin{equation}\label{eq:gamma_B_ineq}
    |\G_B|\leq |\G_{AB}^{(1)}| + |\G_{BC}^{(1)}| + |\g|.
\end{equation}
Furthermore, $X^\prime_{ABC}:=E_{AB}\cup E_{BC}$ also satisfies the homology condition
\begin{equation}
    \partial E^\prime_{ABC} = ABC\cup (\G_{AB}^{(2)}\cup \G_{BC}^{(2)}\cup \g)\cup (Q_{AB}\cup Q_{BC}),
\end{equation}
thus we have
\begin{equation}\label{eq:gamma_ABC_ineq}
    |\G_{ABC}|\leq |\G_{AB}^{(2)}| + |\G_{BC}^{(2)}| + |\g|.
\end{equation}

Finally, by adding the two inequalities \eqref{eq:gamma_B_ineq} and \eqref{eq:gamma_ABC_ineq}, we have
\begin{align}
    |\G_B| + |\G_{ABC}| &\leq
    (|\G_{AB}^{(1)}| + |\G_{AB}^{(2)}|+|\g|) \\&\quad + (|\G_{BC}^{(1)}| + |\G_{BC}^{(2)}|+|\g|)\\
    &=|\G_{AB}| + |\G_{BC}|.
\end{align}
Therefore, SSA on $\Sigma$ holds.

It is also possible to show that the monogamy of mutual information (MMI) \cite{Hayden:2011ag} holds for the same setup of AdS/BCFT at a time slice.
MMI states that for any choice of subsystems, the following inequality holds:
\be
    \begin{split}
        &I_3(A:B:C) \coloneqq S_A + S_B + S_C \\ &\qquad - S_{AB} - S_{BC} - S_{CA} + S_{ABC} \leq 0
    \end{split}
\ee
MMI has been shown to hold in AdS/CFT setups without EOW branes in~\cite{Hayden:2011ag}.
Appendix A provides the details of the proof of MMI in the case of static AdS/BCFT using similar arguments to that of SSA.\\

{\bf Acknowledgements}

We are very grateful to Horacio Casini for valuable comments on the draft of this article. We also thank Isaac Kim, Yuya Kusuki and Masamichi Miyaji for useful discussions.
This work is supported by MEXT KAKENHI Grant-in-Aid for Transformative Research Areas (A) through the ``Extreme Universe'' collaboration: Grant Number 21H05187.
TT is also supported by Inamori Research Institute for Science and World Premier International Research Center Initiative (WPI Initiative)
from the Japan Ministry of Education, Culture, Sports, Science and Technology (MEXT),
by JSPS Grant-in-Aid for Scientific Research (A) No.~21H04469.

\appendix

\section{Appendix A: SSA and MMI in AdS/BCFT}
In this section, we show that SSA and MMI on a time slice hold for any static setups of AdS$_3$/BCFT$_2$.
SSA states that for any three intervals $A$, $B$ and $C$ on a time slice, the following inequality holds,
\begin{equation}
    S_{ABC} + S_{B} \leq S_{AB} + S_{BC}.
\end{equation}
The tripartite mutual information,
\begin{equation}
    \begin{split}
        &I_3(A:B:C) \coloneqq S_A + S_B + S_C \\ &\qquad - S_{AB} - S_{BC} - S_{CA} + S_{ABC},
    \end{split}
\end{equation}
can be either positive or negative in general quantum systems.
MMI states that $I_3(A:B:C) \leq 0$ holds for any choice of subsystems.

We consider a set $\mathcal{A} = \{A, B, C\}$ consisting of three intervals on a time slice with an EOW brane $Q$.
The holographic entanglement entropy of a subsystem $X \in 2^{\mathcal{A}}$ is given by the area of the corresponding extremal surface $\Gamma_X$ in the bulk spacetime,
\begin{equation}
    S_X = \frac{|\Gamma_X|}{4G_N},
\end{equation}
where $G_N$ is the Newton's constant.
The extremal surface $\Gamma_X$ and the EOW brane surface $Q_X \subset Q$ are taken so that $\Gamma_X \cup Q_X$ is homologous to the subsystem $X$.
In other words, there exists a corresponding homology region $E_X$ such that it is enclosed as
\begin{equation}
    \partial(E_X) = X \cup \Gamma_X \cup Q_X.
\end{equation}

Recalling that the entanglement wedge preserves the order of inclusion, we have the following relations,
\begin{equation}
    E_B \subset E_{X} \subset E_{ABC} \quad \text{for} \quad X = AB, BC.
\end{equation}
This implies that the intersection $E_{AB} \cap E_{BC}$ is non-empty.
By cyclic permutation of $A$, $B$ and $C$, we also have the similar relations for the intersection $E_{CA} \cap E_{AB}$ and $E_{BC} \cap E_{CA}$.
Therefore, the extremal surface $\Gamma_{AB}$ is divided into four parts, 
\begin{equation}
    \Gamma_{AB} = \Gamma_{AB}^{(1)} \sqcup \Gamma_{AB}^{(2)} \sqcup \Gamma_{AB}^{(3)} \sqcup \Gamma_{AB}^{(4)},
\end{equation}
where
\begin{align}
    \Gamma_{AB}^{(1)} &\coloneqq \Gamma_{AB} \cap E_{BC} \cap E_{CA} \\
    \Gamma_{AB}^{(2)} &\coloneqq \Gamma_{AB} \cap E_{BC} \setminus E_{CA} \\
    \Gamma_{AB}^{(3)} &\coloneqq \Gamma_{AB} \cap E_{CA} \setminus E_{BC} \\
    \Gamma_{AB}^{(4)} &\coloneqq \Gamma_{AB} \setminus E_{BC} \setminus E_{CA}.
\end{align}
Similarly, we define $\Gamma_{BC}^{(i)}$ and $\Gamma_{CA}^{(i)}$ for $i=1, 2, 3, 4$ by cyclic permutations of $A$, $B$ and $C$.
We also define $Q_{BC}^{(i)}$ and $Q_{CA}^{(i)}$ for $i=1, 2, 3, 4$ in the same manner.

\subsection{Proof of SSA on a time slice}
The bulk region $E_B' \coloneqq E_{AB} \cap E_{BC}$ satisfies the homology condition $\partial(E_B') = B \cup \Gamma_B' \cup Q_B'$ with respect to $B$, where we define
\begin{align}
    \Gamma_B' &\coloneqq \Gamma_{AB}^{(1)} \cup \Gamma_{AB}^{(2)} \cup \Gamma_{BC}^{(3)} \cup \Gamma_{BC}^{(1)}, \\
    Q_B' &\coloneqq Q_{AB}^{(1)} \cup Q_{AB}^{(2)} \cup Q_{BC}^{(3)} \cup Q_{BC}^{(1)} .
\end{align}
By the extremality of $\Gamma_B$, we have the condition
\begin{equation}
    |\Gamma_B| \leq |\Gamma_B'| \leq |\Gamma_{AB}^{(1)}| + |\Gamma_{AB}^{(2)}| + |\Gamma_{BC}^{(3)}| + |\Gamma_{BC}^{(1)}|.
\end{equation}
the area of $\Gamma_B'$ is minimized among all the surfaces homologous to $B$.

The bulk region $E_{ABC}' \coloneqq E_{AB} \cup E_{BC}$ satisfies the homology condition $\partial(E_{ABC}') = (A \cup B \cup C )\cup \Gamma_{ABC}' \cup Q_{ABC}'$ with respect to $A \cup B \cup C$, where we define
\begin{align}
    \Gamma_{ABC}' &\coloneqq \Gamma_{AB}^{(3)} \cup \Gamma_{AB}^{(4)} \cup \Gamma_{BC}^{(4)} \cup \Gamma_{BC}^{(2)}, \\
    Q_{ABC}' &\coloneqq Q_{AB}^{(3)} \cup Q_{AB}^{(4)} \cup Q_{BC}^{(4)} \cup Q_{BC}^{(2)}.
\end{align}
By the extremality of $\Gamma_{ABC}$, we have the condition
\begin{equation}
    |\Gamma_{ABC}| \leq |\Gamma_{ABC}'| \leq |\Gamma_{AB}^{(3)}| + |\Gamma_{AB}^{(4)}| + |\Gamma_{BC}^{(4)}| + |\Gamma_{BC}^{(2)}|.
\end{equation}

Thus, we have
\begin{align}
    & |\Gamma_B| + |\Gamma_{ABC}| \notag\\
    & \quad \leq \sum_{i = 1}^4 \left( |\Gamma_{AB}^{(i)}| + |\Gamma_{BC}^{(i)}| \right) = |\Gamma_{AB}| + |\Gamma_{BC}|,
\end{align}
which implies the SSA.

\subsection{Proof of MMI on a time slice}
The bulk region $\tilde{E}_B \coloneqq E_{AB} \cap E_{BC} \setminus E_{CA}$ satisfies the homology condition $\partial(\tilde{E}_B) = B \cup \tilde{\Gamma}_B \cup \tilde{Q}_B$ with respect to $B$, where we define
\begin{align}
    \tilde{\Gamma}_B &\coloneqq \Gamma_{CA}^{(1)} \cup \Gamma_{AB}^{(2)} \cup \Gamma_{BC}^{(3)}, \\
    \tilde{Q}_B &\coloneqq Q_{CA}^{(1)} \cup Q_{AB}^{(2)} \cup Q_{BC}^{(3)}.
\end{align}
By the extremality of $\Gamma_B$, we have the condition
\begin{equation}
    |\Gamma_B| \leq |\tilde{\Gamma}_B| \leq |\Gamma_{CA}^{(1)}| + |\Gamma_{AB}^{(2)}| + |\Gamma_{BC}^{(3)}|.
\end{equation}
We obtain the similar condition for $E_C$ and $E_A$ as
\begin{align}
    |\Gamma_C| &\leq |\tilde{\Gamma}_C| \leq |\Gamma_{AB}^{(1)}| + |\Gamma_{BC}^{(2)}| + |\Gamma_{CA}^{(3)}|, \\
    |\Gamma_A| &\leq |\tilde{\Gamma}_A| \leq |\Gamma_{BC}^{(1)}| + |\Gamma_{CA}^{(2)}| + |\Gamma_{AB}^{(3)}|
\end{align}
by cyclic permutations of $A$, $B$ and $C$.

The bulk region $\tilde{E}_{ABC} \coloneqq E_{AB} \cup E_{BC} \cup E_{CA}$ satisfies the homology condition $\partial(\tilde{E}_{ABC}) = (A \cup B \cup C )\cup \tilde{\Gamma}_{ABC} \cup \tilde{Q}_{ABC}$ with respect to $A \cup B \cup C$, where we define
\begin{align}
    \tilde{\Gamma}_{ABC} &\coloneqq \Gamma_{AB}^{(4)} \cup \Gamma_{BC}^{(4)} \cup \Gamma_{CA}^{(4)}, \\
    \tilde{Q}_{ABC} &\coloneqq Q_{AB}^{(4)} \cup Q_{BC}^{(4)} \cup Q_{CA}^{(4)}.
\end{align}
By the extremality of $\Gamma_{ABC}$, we have the condition
\begin{equation}
    |\Gamma_{ABC}| \leq |\tilde{\Gamma}_{ABC}| \leq |\Gamma_{AB}^{(4)}| + |\Gamma_{BC}^{(4)}| + |\Gamma_{CA}^{(4)}|.
\end{equation}

Thus, we have
\begin{align}
    &|\Gamma_A| + |\Gamma_B| + |\Gamma_C| + |\Gamma_{ABC}| \notag\\
    & \quad \leq \sum_{i = 1}^4 \left( |\Gamma_{AB}^{(i)}| + |\Gamma_{BC}^{(i)}| + |\Gamma_{CA}^{(i)}| \right) \notag\\
    & \quad = |\Gamma_{AB}| + |\Gamma_{BC}| + |\Gamma_{CA}|,
\end{align}
which implies the MMI.

\bibliography{BCFTSSA.bib}


\end{document}